\documentclass[letter,twocolumn,showpacs,aps,prb,floatfix]{revtex4}
\usepackage{graphicx,subfigure,epsfig,verbatim,psfrag,amsmath,amssymb,color}
\newcommand{\red}[1]{{\textcolor{red}{#1}}}
\newcommand{\blue}[1]{{\textcolor{blue}{#1}}}
\newcommand{\mean}[1]{\left<#1\right>}

\makeatletter
\input epsf

\def\be{\begin{equation}}
\def\ee{\end{equation}}
\def\ba{\begin{eqnarray}}
\def\ea{\end{eqnarray}}
\makeatother

\newcommand{\up}{\uparrow}
\newcommand{\dn}{\downarrow}

\newcommand{\hatSS}{\hat{\mathbf{S}}}

\begin{document}

\title{Dimers on the Triangular Kagome Lattice}
\author{Y.~L.~Loh}
\author{Dao-Xin Yao}
\author{E.~W.~Carlson}
\affiliation{Department of Physics, Purdue University, West Lafayette, IN 47907}
\date{\today}

\begin{abstract}
We derive exact results for close-packed dimers on the triangular
kagome lattice (TKL), formed by inserting triangles into the triangles
of the kagome lattice.  Because the TKL is a non-bipartite lattice,
dimer-dimer correlations are short-ranged, so that the ground state at
the Rokhsar-Kivelson (RK) point of the corresponding quantum dimer
model on the same lattice is a short-ranged spin liquid.  Using the
Pfaffian method, we derive an exact form for the free energy, and we
find that the entropy is $\frac{1}{3} \ln 2$ per site, regardless of
the weights of the bonds.  The occupation probability of every bond is
$\frac{1}{4}$ in the case of equal weights on every bond.  Similar to
the case of lattices formed by corner-sharing triangles (such as the
kagome and squagome lattices), we find that the dimer-dimer
correlation function is identically zero beyond a certain (short)
distance. We find in addition that monomers are deconfined on the TKL,
indicating that there is a short-ranged spin liquid phase at the RK point.
We also find exact results for the ground state energy of
the classical Heisenberg model.  The ground state can be
ferromagnetic, ferrimagnetic, locally coplanar, or locally canted,
depending on the couplings.  From the dimer model and the classical
spin model, we derive upper bounds on the ground state energy of the
quantum Heisenberg model on the TKL.
\end{abstract}
\maketitle

\section{Introduction}

The nontrivial statistical
mechanics problem of dimer coverings of lattices,
which may be used to model, {\em e.g.},
the adsorption of diatomic molecules onto a surface\cite{diatomic},
experienced a  renaissance 
with the discovery of exact mappings to Ising models\cite{fisher-1961,kasteleyn1963}.
A second renaissance came with 
the search for\cite{anderson-rvb-1973,rokhsar-kivelson-1988}
and discovery of\cite{sondhi-triangular}
a true spin liquid phase with deconfined spinons.
In the latter case, the problem of classical dimer coverings 
of a lattice illuminates the physics of the corresponding 
{\em quantum} dimer model.  
At the Rokhsar-Kivelson (RK) point of the quantum dimer model,
the ground states are an equal amplitude superposition of 
dimer coverings within the same 
topological sector,\cite{rokhsar-kivelson-1988,sondhi-triangular}\footnote{A 
topological sector is defined as the following. Draw a line through the system, without touching any site. For a given topological sector, the number of dimers which cross that line is invariant modulo 2 under local rearrangements of the dimer covering.  See, {\em e.g.}, Ref.~\onlinecite{sachdevRMP}.}
and in fact dimer correlations at this point correspond
to the dimer correlations of the classical dimer model. 

Results on classical hard core dimer models in two\cite{kasteleyn1963} 
and higher dimensions\cite{huse-prl-2003} point to two classes of models,
depending upon the monomer-monomer correlation function, 
which is defined as the ratio of the number of configurations available
with two test monomers inserted to the number of configurations
available with no monomers present.
On bipartite lattices (such as the square and honeycomb lattices),
monomers are confined, with power law correlations.\cite{kasteleyn1963,fisher1963}
On nonbipartite lattices (such as the triangular, kagome,
and the triangular kagome lattice discussed here), 
monomers can be either confined or deconfined,
and correlators exhibit exponential decay 
except at phase transtions.\cite{krauth-2003,fendley-2002,misguich-2002,fisher1966,moessner-2003}
This implies that 
while the RK point of the quantum dimer model is critical on bipartite lattices,
so that at T=0 a (critical) spin liquid exists only at the RK point,
in non-bipartite lattices, such as the triangular lattice
and lattices made of corner-sharing triangles such as
the kagome and squagome lattice, it has been shown
that the RK point corresponds to a disordered spin liquid.
Correspondingly, it was established in both of these cases
that there exist finite regions of parameter space where
the ground state is a gapped spin liquid with deconfined spinons.
Part of the interest in such states is the topological
order that accompanies such ground states, and hence
such states may be useful examples of the toric code.
Interest also stems from the original proposals that the
{\em doped} spin liquid phase leads to superconductivity.\cite{anderson-1987,rokhsar-kivelson-1988}

\begin{figure}[tb]
\includegraphics[width=0.85\columnwidth]{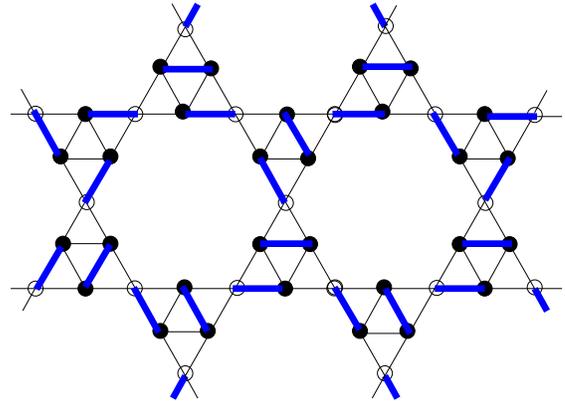}
\caption{(Color online) A dimer covering of a portion of the 
triangular kagome lattice (TKL).
The TKL can 
be derived from the triangular lattice by periodically deleting seven out of
every sixteen lattice sites.  
This structure has two different sublattices
``a'' (closed circles) and ``b'' (open circles), 
which correspond to small trimers and large trimers,
respectively.  
Each site has four nearest neighbors.  The primitive unit cell contains 6 $a$-sites, 3 $b$-sites, 6 $a$--$a$ bonds, and 12 $a$--$b$ bonds.  Thick blue lines represent dimers.  A typical close-packed dimer
covering is shown.
}
\label{f:valence-bonds}
\end{figure}

In this paper, we analyze the problem of classical close-packed dimers
on the triangular kagome lattice (TKL), a non-bipartite lattice
expected to display a spin liquid phase, as the first step in
understanding the RK point of the corresponding quantum dimer model.
The TKL, depicted in Fig.~\ref{f:valence-bonds}, has a 
physical analogue in the positions of Cu atoms in the materials
$\mbox{Cu}_{9}\mbox{X}_2(\mbox{cpa})_{6}\cdot x\mbox{H}_2\mbox{O}$
(cpa=2-carboxypentonic acid, a derivative of ascorbic acid; X=F,Cl,Br)
~\cite{gonzalez93,maruti94, mekata98}.  We have previously studied
Ising spins\cite{lohyaocarlson2007} and XXZ/Ising spins\cite{xxzising}
on the TKL; this paper represents an alternative approach to the
problem.  Using the well-known Pfaffian method,\cite{kasteleyn1963} we
obtain exact solutions of close-packed dimers on the TKL.  We obtain
an analytic form of the free energy for arbitrary bond weights.  The
entropy is $\frac{1}{3} \ln 2$ per site, independent of the weights of
the bonds, $z_{aa}$ and $z_{ab}$.  We find the occupation probability
of every bond is a constant $\frac{1}{4}$ in the absence of an
orienting potential.  The system has only local correlations, in that
the dimer-dimer correlation function is exactly zero beyond two
lattice constants, much like the situation on lattices made from
corner-sharing triangles such as the kagome and squagome
lattices\cite{misguich-2002}.
We use exact methods to find the monomer-monomer correlation function,
and show that monomers are deconfined on the TKL.  In addition, we
solve for the ground states of the classical Heisenberg on this model.
In addition to collinear phases (ferromagnetic and ferrimagnetic), we
find a canted ferrimagnetic phase which interpolates smoothly between
the two.  We obtain a variational upper bound to the ground state energy
of the TKL quantum Heisenberg antiferromagnet using closed-packed
dimer picture. 


\section{Model, Thermodynamic Properties, and Correlation Function \label{model}}
\begin{figure}[t]
\psfrag{z1}{\Large $z_1$} \psfrag{z2}{\Large $z_2$} \psfrag{z3}{\Large $z_3$}
\psfrag{z4}{\Large $z_4$} \psfrag{z5}{\Large $z_5$} \psfrag{z6}{\Large $z_6$}
\includegraphics[width=0.7\columnwidth]{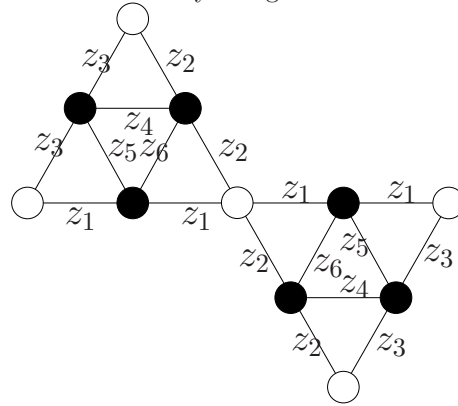}
\caption{Our assignment of weights $z_{\alpha}$ to bonds in the TKL.  Solid (open) circles represent $a$-sites ($b$-sites).
\label{f:weights}}
\end{figure}

In this paper we consider the close-packed dimer model on the TKL,
a lattice which can be obtained by inserting triangles inside of the triangles
of the kagome lattice (see Fig.~\ref{f:valence-bonds}). 
The dimer generating function is defined as  
        \begin{align}
        Z &= \sum_{\text{dimer coverings}}  \prod_{\mean{ij}} z_{ij}  {} ^ {n^{ij}} ,
        \label{e:partition-function}
        \end{align}
where $\mean{ij}$ indicates a product over nearest-neighbor bonds, $z_{ij}$ is the weight on the bond joining site $i$ and site $j$, and $n_{ij}$ is the number of dimers (either 0 or 1) on bond $ij$ for the dimer covering under consideration.  The term ``close-packed'' refers to the constraint
that every lattice site must be occupied by one dimer, that is, that vacancies are not
allowed.  Therefore the number of sites $N_\text{sites}$ is twice the number of dimers $N_\text{dimers}=\sum_{\mean{ij}} n_{ij}$.
We allow for the possibility of different weights $z_{\alpha}=e^{-\beta\epsilon_{\alpha}}$ for six different types of bonds $\alpha=1,2,3,4,5,6$, as depicted in Fig.~\ref{f:weights}.  
Figure~\ref{f:valence-bonds} shows an example of a dimer covering.

\begin{figure}[htb]
\includegraphics[width=0.9\columnwidth]{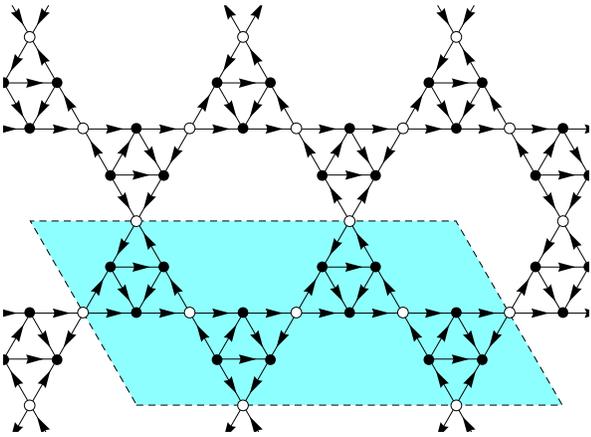}
\caption{(Color online) The arrows represent a Kasteleyn orientation on the triangular kagome lattice (TKL).  Solid (open) circles represent ``a'' (``b'') sublattices.  The shaded region represents the doubled unit cell.
\label{pfaffian-orientation}}
\end{figure}

Several properties of this model, including
the free energy, entropy, and dimer-dimer correlation function, can be 
calculated exactly using the well-known Pfaffian method\cite{kasteleyn1963}.  
We begin by defining a Kasteleyn orientation\cite{kasteleyn1963} (or Pfaffian orientation) for this lattice,
{\em i.e.} a pattern of arrows laid on the bonds such
that in going clockwise around any closed loop with an even number of bonds,
there is an odd number of arrows pointing in the clockwise direction
along the bonds. 
For the TKL, we have found it necessary to double 
the unit cell in order to obtain a valid Kasteleyn orientation.  \footnote{Kasteleyn's theorem
  may be generalized to allow complex phase factors in the weighted adjacency
  matrix: for a transition cycle passing through sites $1,2,\dotsc,2n$, the
  phase factors must satisfy $\eta_{12} \eta_{34} \dotso \eta_{2n-1,2n} =
  -\eta_{23} \eta_{45} \dotso \eta_{2n,1}$.  Complex phase factors provide a
  more elegant solution of the square lattice dimer model\cite{wu1962}.  However, they do not help in the case of the kagome lattice\cite{wang:040105} or TKL; we have found
  that any orientation with the periodicity of the original lattice violates
  the generalized Kasteleyn theorem, even if the phase factors are allowed to
  be arbitrary complex numbers.
  }  
 Such an orientation is shown in
Fig.~\ref{pfaffian-orientation}.  The doubled unit cell contains 18 sites.

The antisymmetric weighted adjacency matrix associated with this
orientation, $A_{ij}$, is a $N_\text{sites} \times N_\text{sites}$ square matrix with a ``doubly Toeplitz'' block structure.   The generating function of the dimer model is given by the Pfaffian of this matrix: $Z = \text{Pf}~ \mathbf{A} = \sqrt{ \det \mathbf{A}}$.  In the infinite-size limit, this approaches an integral over the two-dimensional Brillouin zone:
\begin{eqnarray}
f &=& \lim_{N_\text{sites} \rightarrow \infty} \frac{F}{N_\text{sites}} \\  \nonumber
&=& \frac{1}{18}
        \int_0^{2\pi} \frac{dk_x}{2\pi}
        \int_0^{2\pi} \frac{dk_y}{2\pi}
        ~\tfrac{1}{2} \ln \left|  \det \mathbf{M} (k_x, k_y)   \right| 
\end{eqnarray}
where  we have normalized the free energy by the temperature such that $F \equiv {\rm ln}Z$, 
and where  $\mathbf{M}(k_x,k_y)$ is the 18x18 matrix below,
%
%
\begin{widetext}
\begin{align}
\mathbf{M} &= 
\left(
\begin{array}{llllllllllllllllll}
  0 & z_1 & z_3 & 0 & 0 & 0 & 0 & 0 & 0 & 0 & 0 & 0 & 0 & 0 & 0 & 0 & \frac{z_3}{u} & -\frac{z_1}{u} \\
 -z_1 & 0 & -z_5 & 0 & 0 & z_1 & -z_6 & 0 & 0 & 0 & 0 & 0 & 0 & 0 & 0 & 0 & 0 & 0 \\
 -z_3 & z_5 & 0 & -v z_3 & 0 & 0 & z_4 & 0 & 0 & 0 & 0 & 0 & 0 & 0 & 0 & 0 & 0 & 0 \\
 0 & 0 & \frac{z_3}{v} & 0 & -z_2 & 0 & -\frac{z_2}{v} & -z_3 & 0 & 0 & 0 & 0 & 0 & 0 & 0 & 0 & 0 & 0 \\
 0 & 0 & 0 & z_2 & 0 & z_2 & 0 & z_4 & z_6 & 0 & 0 & 0 & 0 & 0 & 0 & 0 & 0 & 0 \\
 0 & -z_1 & 0 & 0 & -z_2 & 0 & z_2 & 0 & z_1 & 0 & 0 & 0 & 0 & 0 & 0 & 0 & 0 & 0 \\
 0 & z_6 & -z_4 & v z_2 & 0 & -z_2 & 0 & 0 & 0 & 0 & 0 & 0 & 0 & 0 & 0 & 0 & 0 & 0 \\
 0 & 0 & 0 & z_3 & -z_4 & 0 & 0 & 0 & -z_5 & -z_3 & 0 & 0 & 0 & 0 & 0 & 0 & 0 & 0 \\
 0 & 0 & 0 & 0 & -z_6 & -z_1 & 0 & z_5 & 0 & z_1 & 0 & 0 & 0 & 0 & 0 & 0 & 0 & 0 \\
 0 & 0 & 0 & 0 & 0 & 0 & 0 & z_3 & -z_1 & 0 & z_1 & z_3 & 0 & 0 & 0 & 0 & 0 & 0 \\
 0 & 0 & 0 & 0 & 0 & 0 & 0 & 0 & 0 & -z_1 & 0 & -z_5 & 0 & 0 & z_1 & -z_6 & 0 & 0 \\
 0 & 0 & 0 & 0 & 0 & 0 & 0 & 0 & 0 & -z_3 & z_5 & 0 & -v z_3 & 0 & 0 & z_4 & 0 & 0 \\
 0 & 0 & 0 & 0 & 0 & 0 & 0 & 0 & 0 & 0 & 0 & \frac{z_3}{v} & 0 & z_2 & 0 & -\frac{z_2}{v} & z_3 & 0 \\
 0 & 0 & 0 & 0 & 0 & 0 & 0 & 0 & 0 & 0 & 0 & 0 & -z_2 & 0 & z_2 & 0 & z_4 & z_6 \\
 0 & 0 & 0 & 0 & 0 & 0 & 0 & 0 & 0 & 0 & -z_1 & 0 & 0 & -z_2 & 0 & z_2 & 0 & z_1 \\
 0 & 0 & 0 & 0 & 0 & 0 & 0 & 0 & 0 & 0 & z_6 & -z_4 & v z_2 & 0 & -z_2 & 0 & 0 & 0 \\
 -u z_3 & 0 & 0 & 0 & 0 & 0 & 0 & 0 & 0 & 0 & 0 & 0 & -z_3 & -z_4 & 0 & 0 & 0 & -z_5 \\
 u z_1 & 0 & 0 & 0 & 0 & 0 & 0 & 0 & 0 & 0 & 0 & 0 & 0 & -z_6 & -z_1 & 0 & z_5 & 0
\end{array}
\right)
\end{align}
\end{widetext}
where, for brevity, we have written $u=e^{ik_x}$ and $v=e^{ik_y}$.
The determinant of this matrix is independent of $k_x$ and $k_y$:
\begin{align}
&\det \mathbf{M} (k_x, k_y) 
\nonumber\\
&=
64 z_1^2 z_2^2 z_3^2 \left(z_1 z_4+z_2 z_5\right){}^2 \left(z_1 z_4+z_3 z_6\right){}^2 \left(z_2 z_5+z_3 z_6\right){}^2
\end{align}
Taking the logarithm and integrating over the Brillouin zone gives the free energy per doubled unit cell.
Hence, the free energy per site  is
\begin{eqnarray}
f &=&\frac{1}{18} \ln\big[ 8 z_1 z_2 z_3 \left(z_1 z_4+z_2 z_5\right)  \\ \nonumber
& &\hspace{.25in}\times \left(z_1 z_4+z_3 z_6\right) \left(z_2 z_5+z_3 z_6\right)\big]~. 
\label{e:free-energy}
\end{eqnarray}

The occupation probability of each bond may be calculated by
differentiating the free energy with respect to the weight of each bond.
Let $N_\alpha$ be the total number of dimers on $z_\alpha$-bonds (as defined in Fig.~\ref{f:weights}), averaged over all configurations of the system.  Since $Z=\sum_\text{configs} \prod_\alpha z_\alpha {}^ {N_\alpha}$, we have
$N_\alpha = z_\alpha \frac{\partial F}{\partial z_\alpha}$.  
We define the occupation probability of each $\alpha$-bond  as $p_\alpha = \frac{N_\alpha}{B_\alpha}$, where $B_\alpha$ is the total number of type-$\alpha$ bonds on the lattice.  If $N_\text{cells}$ is the number of primitive unit cells, then $N_\text{sites}=9N_\text{cells}$, $B_1=B_2=B_3=4N_\text{cells}$, and $B_4=B_5=B_6=2N_\text{cells}$.  
The results, normalized by the number of sites in the system, are
\begin{align}
p_1 &= \frac{1}{8}
\left(1 + \frac{z_1 z_4 }{z_1 z_4+z_3 z_6}+\frac{z_1 z_4 }{z_1 z_4+z_2 z_5}\right).\\
p_4 &= \frac{1}{4}
\left(\frac{z_1 z_4 }{z_1 z_4+z_3 z_6}+\frac{z_1 z_4 }{z_1 z_4+z_2 z_5}\right),
\end{align}
Expressions for $p_2$, $p_3$, $p_5$, $p_6$
follow by cyclic permutation of $\{1,2,3\}$ 
simultaneously with permutation of $\{4,5,6\}$.
The entropy can be computed by
the usual Legendre transformation, 
$S=F + \sum_{\alpha=1}^6 \beta\epsilon_\alpha N_\alpha$.  
\footnote{The expression for $S$ does not simplify appreciably
when $F$, etc., are substituted in.}

The behavior of the correlation functions can be deduced in the same way as in Ref.~\onlinecite{misguich-2002}.  
To find the dimer-dimer correlation functions, the standard method is to
first calculate the ``fermion'' Green function, which is the inverse of the
matrix $\mathbf{A}$, Fourier-transform it to real space, and use the result
to construct the dimer-dimer correlation functions.  The inverse of the
matrix $\mathbf{A}$, $\mathbf{G}(k_x,k_y)= \left[ \mathbf{A}(k_x,k_y)
\right]^{-1}$, can be written as the matrix of cofactors of $\mathbf{A}$
divided by the determinant of $\mathbf{A}$.  Since $\det\mathbf{A}$ is
independent of $k_x$ and $k_y$, the only dependence on $k_x$ and $k_y$ enters
through the cofactor matrix.  Each cofactor is at most a monomial in $e^{ik_x}$ and
$e^{ik_y}$.  From the rules of Fourier transformation it is easily seen that
the real-space Green function $\mathbf{G}(x,y)$ is zero when $|x|>1$ or $|y|>1$
is greater than a certain cutoff distance.
Hence the dimer-dimer correlation function will be zero beyond a distance of two unit cells.
This is true \emph{regardless of the values of the bond weights} depicted in Fig.~\ref{f:weights}.
This extremely short-ranged behavior of the correlation function is
similar to that for dimers on the kagome lattice\cite{misguich-2002},
and also to the spin-spin correlation for Ising spins in
the frustrated parameter regime.~\cite{lohyaocarlson2007}
It underscores the special role played by kagome-like lattices (c.f. Ref.~\onlinecite{wang:040105,misguich-2002}).

Whereas quantum dimer models on bipartite lattices do not support deconfined spinons,
quantum dimer models on non-bipartite lattices can have deconfined spinons.
The connection to classical dimer models is that at the RK point,
correlatons in the quantum dimer model are the same as the correlations
of the corresponding classical dimer problem.
The only non-bipartite lattice for which deconfined spinons have been
rigorously demonstrated is the triangular lattice, by explicitly
calculating the classical monomer-monomer correlation function
using Pfaffian methods.\cite{fendley-2002}
On the kagome lattice, while no correspondingly rigorous calculation
of the monomer-monomer correlation function has yet been demonstrated,
there have been several indications that the spinons in quantum dimer models
on the kagome lattice are deconfined (and therefore classical monomer-monomer
correlators are similarly deconfined), from {\em e.g.}, 
the energetics of static spinon configurations\cite{misguich-2002},
the behavior of the single-hole spectral function\cite{poilblanc-2004},
and in the limit of easy-axis anisotropy\cite{balents-2002}.
We have calculated the monomer-monomer correlation for the kagome lattice dimer
model using the Pfaffian approach of Fisher and Stephenson\cite{fisher1963} and we find that it is
strictly constant, with $M(r)=1/4$ for any $r>0$.
\footnote{The long distance behavior can also be seen from the perspective of the
quantum dimer model of Ref.~\onlinecite{misguich-2002},
in that each monomer removes an Ising degree of freedom, since
it merges two hexagons.  Therefore each monomer removes
half of the configurations.\cite{misguich-private}}
Because the triangular kagome lattice dimer model
maps to the kagome dimer model (with an extra degeneracy of 4 per unit
cell), the monomer-monomer correlation on the TKL is also $M(r)=1/4$
for monomers on any two $b$-sites, or for any combination of $a$ and $b$ 
sites at least three sites apart.


\section{Effects of an Orienting Potential}

In the $\mbox{Cu}_{9}\mbox{X}_2(\mbox{cpa})_{6}\cdot
x\mbox{H}_2\mbox{O}$ materials\cite{gonzalez93,maruti94, mekata98}, 
the $a$ spins are closer to each other than they are to the $b$ spins, so the exchange couplings satisfy $|J_{aa}| > |J_{ab}|$.  
In the classical dimer approximation described in Sec.~\ref{s:heisenberg}, this corresponds to
unequal weights for dimers on $ab$ bonds {\em vs.} those on $aa$ bonds, 
 $|z_{aa}| > |z_{ab}|$.    Aside from this intrinsic difference in bond weights, 
it may also be possible to apply anisotropic mechanical strain to vary the lattice geometry (and hence the exchange couplings and dimer weights) in different directions.

\begin{figure}[h]
{\centering
  \subfigure[Bond occupation probabilities\label{f:p}]{\includegraphics[width=0.85\columnwidth]{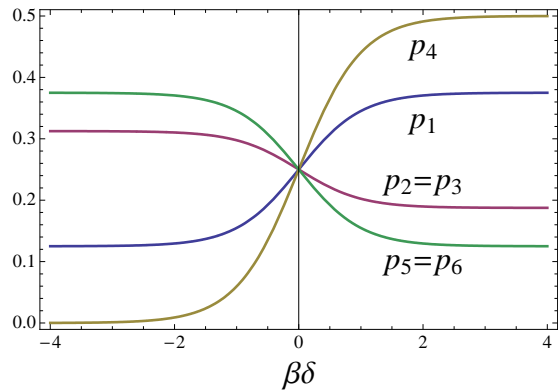}}
  \subfigure[Entropy per site\label{f:s}]{\includegraphics[width=0.85\columnwidth]{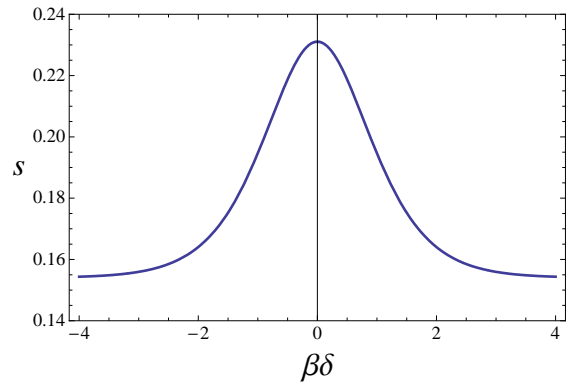}}
}
\caption{(Color online) Bond occupation probabilities and entropy per site, as functions of the orienting field $\beta \delta$ defined in the text.
\label{density-and-entropy}}
\end{figure}

To obtain some insight into the behavior of the classical dimer model under these conditions, we write $z_{\alpha}=e^{-\beta\epsilon_{\alpha}}$, where
$\beta=1/T$ is the inverse temperature and $\epsilon_\alpha$ is the potential energy for dimers on bond $\alpha$. 
We use the following parametrization for the potential energy on each site:
        \begin{align}
        \epsilon_1&=\epsilon_{ab} - \delta, \quad 
        \epsilon_2=
        \epsilon_3=\epsilon_{ab}, \\
        \epsilon_4&=\epsilon_{aa} - \delta, \quad 
        \epsilon_5=
        \epsilon_6=\epsilon_{aa},
        \end{align}
where $\delta$ is an orienting potential (i.e., an anisotropy parameter) which favors dimers in one direction.
The bond occupation probabilities and entropy are
independent of the values of $\epsilon_{ab}$ and $\epsilon_{aa}$, and depend
smoothly upon $\beta\delta$ (see Fig.~\ref{density-and-entropy}): 
        \begin{align}
          p_1 &= \frac{1}{8} \left( 2 + \tanh \beta\delta \right), \label{e:p1} \\
          p_4 &= \frac{1}{4} \left( 1 + \tanh \beta\delta \right), \\
          p_2 &= p_3= \frac{1}{16} \left( 4 - \tanh \beta\delta \right), \\
          p_5 &= p_6= \frac{1}{8} \left( 2 - \tanh \beta\delta \right), \label{e:p4} \\
          s &= \frac{S}{N_\text{sites}} = \frac{1}{18} \left[
                         \ln \left(64\cosh^2\beta\delta\right) 
                        - 2\beta\delta\tanh \beta\delta   
               \right]             .
        \end{align}
These results show that the TKL dimer model has neither a deconfinement transition (as a function of $\epsilon_{ab}-\epsilon_{aa}$) nor a Kasteleyn transition (as a function of $\delta$).
It does, however, have a Curie-like ``polarizability'' with respect to an orienting potential.  
This is contrast to the situation on the kagome lattice,\cite{wang:040105} where the bond occupation probabilities do not depend on the orienting potential.



\section{Results for symmetrical case}
In the absence of the orienting potential (i.e., $\delta=0$), the expressions
for the bond occupation probabilities and entropy become very simple:
\begin{align}
&p_\alpha = \frac{1}{4},  \qquad \alpha=1,2,3,4,5,6, \\
&s = \frac{1}{3} \ln 2.
\end{align}
Note that these quantities are independent of the relative bond weights
$z_{aa}$ and $z_{ab}$.  
The comparison with other lattices in Table~\ref{comparison-table}
shows that the entropy per site for the TKL is 
the same as that for the kagome lattice.  Although
the two lattices are related, this is in fact a coincidence
for the following reason.  
The similarity can be seen by considering the number of
$b$-spins per unit cell which have a dimer that
connects to a different unit cell.  Because there is an 
odd number of sites per unit cell, this number must be odd,
{\em i.e.} either $1$ or $3$.   Since the $b$-spins
themselves form a kagome lattice, the same
is in fact true of the kagome lattice.  
The difference is that for a given pattern of external
dimers connecting to $b$ spins, there is no 
further degeneracy in the kagome case,
whereas for the TKL there are four different
internal dimer patterns corresponding to any
given pattern of external dimers connecting to the $b$ spins.
This means that the TKL has a further $4$-fold degeneracy,
so that the kagome entropy per unit cell of $s_{\rm cell} = \ln 2$
becomes an entropy per unit cell of $s_{\rm cell} = \ln 8 = 3 \ln 2 $
in the TKL.
Since there are $9$ spins per unit cell in the TKL, this yields 
$s=(1/3) \ln 2$ per site.

The total numbers of dimers on $a$--$a$ bonds and on $b$--$b$ bonds are
        \begin{align}
        N_{aa}=\frac{1}{3}N_\text{dimers}, 
        \label{e:Naa}
        \\
        N_{ab}=\frac{2}{3}N_\text{dimers},
        \label{e:Nab}
        \end{align}
where $N_\text{dimers}$ is the total number of dimers and $N_\text{dimers}=\frac{1}{2} N_\text{sites}$.   (Of course, $N_{aa} = N_4+N_5+N_6$ and $N_{ab}=N_1+N_2+N_3$.)
Note that because there are twice as many $a$--$b$ bonds in the lattice as there
are $a$--$a$ bonds, this implies that the dimer density {\em is the same on every bond},
regardless of the weights of the bonds.   
Since the number of sites is twice the number of dimers in the close-packed case,
$N_{\rm sites} = 2 N_{\rm dimers}$, there are on average $9/2$ dimers
per unit cell.  One third of those are on the $aa$ bonds, or $3/2$ per unit
cell.  Since there are six $aa$ bonds per cell, there are $(3/2)/6 = 1/4$
dimers per $aa$ bond.
A similar analysis shows that there are $1/4$ dimers per $ab$ bond.
In other words, there are $1/4$ dimers per bond, regardless
of the relative weight $z_{aa}$ and $z_{ab}$, and 
regardless of whether it is an $aa$ or $ab$ bond.  
Under the constraint of close-packing,
the dimer densities are set by {\em geometry}, rather than by energetics,
similar to case of classical dimers on the kagome lattice\cite{wang:040105,wunderlich,elser-1989,elser-1993},

Our results for close-packed, classical dimers on the TKL
are summarized in Table~\ref{comparison-table},
along with known results for the corresponding properties on
the square, honeycomb, triangular, and kagome lattices.
Notice that the kagome and TKL 
are special in having simple, closed-form expressions for the entropies.
In fact, the entropy per unit cell in each case is the 
logarithm of an integer.  
On triangular lattice as well as on the two bipartite lattices
which are shown in the table
(square and honeycomb), the entropy is not expressible
as the logarithm of an integer.

\begin{widetext}
\begin{center}
\begin{table}[h]
\begin{ruledtabular}
\begin{tabular}{ccccc}
Lattice              &  Entropy &  Dimer corr. & Monomer corr. & Polarizability    \\ \hline
Square\cite{fisher1963}
&  $0.2915609$ &  $r^{-2}$          & $r^{-1/2}$ & finite \\
Honeycomb\cite{moessner2003}
&  $0.161533$ &  $r^{-2}$          & $r^{-1/2}$  & Kasteleyn transition \\
Triangular\cite{fendley-2002}
&  $0.4286$   &  $e^{-r/0.6014}$    & const+$e^{-r/0.6014}$ & finite \\
Kagome\cite{misguich-2002,wang:040105}
                                       &  $\frac{1}{3}\ln 2=0.231049$ &  local  & deconfined & 0 \\
TKL
                                 &  $\frac{1}{3}\ln 2=0.231049$ &  local  &  deconfined & finite  \\
\end{tabular}
\end{ruledtabular}
\caption{Properties of close-packed dimer models on various lattices.  Entropies are quoted per site.
``Local'' means that the correlation function is exactly zero beyond a certain radius -- it has ``finite support''.   
The triangular, kagome, and triangular kagome lattices have deconfined monomers.
The honeycomb dimer model not only has a finite dimer polarizability, but it 
 has a Kasteleyn transition at $\delta=\delta_c$.
The polarizability describes the changes in bond occupation probabilities induced by an orienting potential $\delta$.  
\label{comparison-table}}
\end{table}
\end{center}
\end{widetext}

The square and honeycomb lattices, being bipartite,
admit a mapping to a solid-on-solid model\cite{height-model},
and therefore have power-law correlations for both the
dimer-dimer correlations and the monomer-monomer correlations.
In the corresponding quantum dimer models, these 
lattices do not support deconfined spinons.
As conjectured in Ref.~\onlinecite{fendley-2002},
the non-bipartite lattices have exponential (or faster)
falloff of the dimer-dimer correlations.  
In the triangular, kagome, and TKL lattices, monomers are 
deconfined, which means that
spinons are deconfined in the corresponding 
quantum dimer model at the RK point. 
In fact, Moessner and Sondhi showed that there is
a finite region of parameter space in which
a stable spin liquid {\em phase} is present
on the triangular lattice.\cite{sondhi-triangular}


\section{Bounds on the Ground State Energy of the Quantum Heisenberg Model\label{s:heisenberg}}

It is thought that the materials
$\mbox{Cu}_{9}\mbox{X}_2(\mbox{cpa})_{6}\cdot x\mbox{H}_2\mbox{O}$ can
be described in terms of quantum $S=1/2$ spins on the Cu atoms coupled
by 
superexchange interactions.  
Nearest-neighbor isotropic antiferromagnetic couplings
between $S=1/2$ spins on a 2D lattice with sublattice
structure can lead to N\'eel order.
For example, $2$-sublattice N\'eel order is 
favored on the square lattice, whereas
$3$-sublattice N\'eel order is favored on
the triangular lattice.\cite{huse-elser-1988}
However, on the kagome lattice and the TKL,
quantum fluctuations are much more severe, and there is a
possibility that they may lead to alternative ground states (such as
valence-bond liquids).

\begin{figure}[tb]
\includegraphics[width=1.1\columnwidth]{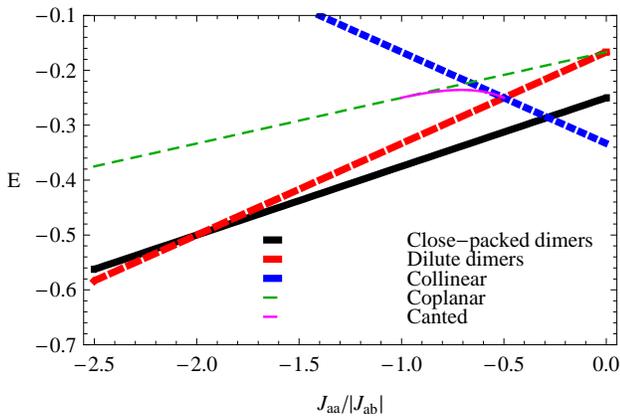}
\caption{(Color online) 
Comparison of upper bounds on the ground state energy per site of the 
quantum Heisenberg model on the TKL, obtained by considering various trial wavefunctions.
In the figure, we have set $S=1/2$. 
}
\label{f:energies}
\end{figure}

A valence bond state is a direct product of singlet pair states.  Using a fermionic representation for the spins,
        \begin{align}
        \left| \Psi_{  \{ n \}  } \right>
        &=
        \Bigg[  
        \prod_{\mean{ij}}
        \frac{1}{\sqrt{2}}
        \left( c^\dag_{i\up} c^\dag_{j\dn} - c^\dag_{i\dn} c^\dag_{j\up} \right)
        ^ {n_{ij}}
        \Bigg]
        \left| \text{vacuum} \right>
        \end{align}
where $n_{ij}=0$ or $1$ is the number of valence bonds on bond $ij$, just as in Eq.~\eqref{e:partition-function}.

Consider a quantum Hamiltonian with isotropic antiferromagnetic Heisenberg
interactions
        \begin{align}
        \hat{H}
        &= -
        \sum_{\mean{ij}}
        J_{ij} \hatSS_i \cdot \hatSS_j,
        \end{align}
where $J_{ij}$ is negative.  
 The expectation value of this Hamiltonian in a valence-bond state is
        \begin{align}
        \left< \Psi_{  \{ n \}  }  \right|  
        J_{ij} \hatSS_i \cdot \hatSS_j
        \left| \Psi_{  \{ n \}  } \right>
        &=
        -\frac{3}{4} \sum_{\mean{ij}} n_{ij} |J_{ij}|~.
        \end{align}
For close-packed dimers, the densities of valence bonds on $aa$ and $ab$ bonds are given by
Eqs.~\eqref{e:Naa} and \eqref{e:Nab}.  Therefore, the total energy of the close-packed valence-bond ``trial wavefunction'' is
        \begin{align} 
        E_\text{VB}&=-\frac{3}{4} \big( N_{aa}|J_{aa}| + N_{ab}|J_{ab}| \big)\\
       &=-\frac{1}{4} \big( |J_{aa}| + 2|J_{ab}| \big) N_\text{dimers}\\
       &=-\frac{1}{8} \big( |J_{aa}| + 2|J_{ab}| \big) N_\text{sites}.
        \label{e:close-packed-vb-energy}
        \end{align}
This serves as an upper bound
of the ground state energy of the quantum Heisenberg model.  
Of course, matrix elements of the Hamiltonian which connect one
dimer covering to another can serve to lower the actual energy even further.  

One may also consider a more dilute dimer state.
For large $|J_{aa}|$, 
one may expect dimers to preferentially occupy
$a-a$ bonds, so that hexamers with three 
$a-b$ bonds are disallowed.
In such a trial dimer state, the 
associated energy is 
\begin{equation}
E_{\rm dilute} = -{1 \over 6} (|J_{aa}| + |J_{ab}|)N_{sites}~.
\label{e:dilute-vb}
\end{equation}
As shown in Fig.~\ref{f:energies}, this upper bound to the ground
state energy is lower than the others for large $|J_{aa}|$. If $J_{ab}$  
is ferromagnetic and $J_{aa}$ is still antiferromagnetic, we expect
another diluted dimer state, where dimers preferentially occupy $a-a$ bonds,
and other spins tend to be aligned (ferromagnetic phase). The corresponding energy is
\begin{equation}
E_{\rm dilute+FM}= - \left({1 \over 6}|J_{aa}| + {1 \over 9}|J_{ab}| \right)N_{sites}~.
\end{equation}

Other bounds can be obtained by considering the \emph{classical} ground states
of the Heisenberg model on the TKL (in which the spins are 3-vectors of magnitude $S=1/2$).  
In the materials of interest, there is not yet consensus whether
the coupling $J_{ab}$ is ferromagnetic or antiferromagnetic.
However, the Hamiltonian of the classical Heisenberg model is invariant under the transformation 
$S_b \rightarrow - S_b$ with $J_{ab} \rightarrow -J_{ab}$, so the thermodynamics are independent of the sign of $J_{ab}$.

\begin{figure}[t]
\psfrag{alpha}{$\blue\alpha$}
\psfrag{beta}{$\red\beta$}
\includegraphics[width=0.6\columnwidth]{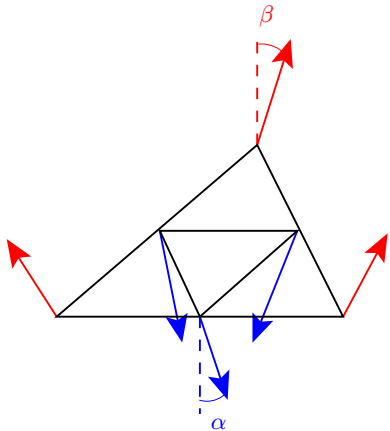}
\caption{(Color online) Canted state of a hexamer of classical Heisenberg spins on the TKL.
$\alpha$ and $\beta$ are the canting angles of the $a$- and $b$-spins from the vertical axis.
When $\alpha=\beta=0$, this reduces to a collinear state (which is ferromagnetic or antiferromagnetic depending on the sign of $J_{ab}$).
When $\alpha=\beta=\pi/2$, it reduces instead to a coplanar state, in which the spins are all at $\pi/3$ to each other. 
}
\label{f:canted}
\end{figure}

\begin{figure}[t]
\includegraphics[width=0.95\columnwidth]{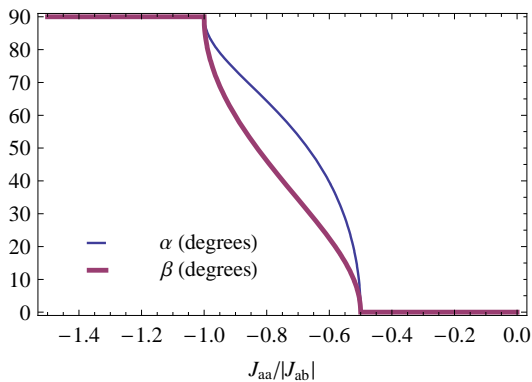}
\caption{(Color online) Canting angles 
in the ground state of the classical Heisenberg model on the TKL
for as a function of coupling ratio $J_{aa}/|J_{ab}|$.
The thin line shows the canting angle $\alpha$ of the $a$-spins,
and the thick line shows the canting angle $\beta$ of the $b$-spins,
with respect to the collinear state, which is ferromagnetic or antiferromagnetic depending on the sign of $J_{ab}$.
}
\label{f:canting-angles}
\end{figure}
First, let us consider classical Heisenberg spins on a \emph{single} hexamer.
By direct minimization of the energy of a single hexamer,
we find that its classical ground state may be either collinear, coplanar, or canted.  
For $J_{aa} > -|J_{ab}|/2$, the ground state is collinear; the $a$-spins are aligned with each other, the $b$-spins are aligned with each other, and the $a$- and $b$-spins are parallel if $J_{ab}$ is ferromagnetic or antiparallel if $J_{ab}$ is antiferromagnetic.   For $J_{aa}<-|J_{ab}|$, the ground state is coplanar; the $a$-spins are at $120^\circ$ to each other, the $b$-spins are at $120^\circ$ to each other, and adjacent $a$- and $b$-spins are at $60^\circ$ if $J_{ab}$ is ferromagnetic or at $120^\circ$ if $J_{ab}$ is antiferromagnetic.   At intermediate couplings, $-|J_{ab}| < J_{aa} < -|J_{ab}|/2$, the ground state is a canted state in which neither the $a$-spins nor the $b$-spins are coplanar; rather, each sublattice is canted away from N\'eel order,
and each sublattice is canted away from the other. 
We define the canting angles of the $a$- and $b$-spins, $\alpha$ and $\beta$, such that
$\alpha = \beta = 0$ in the collinear state (see Fig.~\ref{f:canted}).
The canting angles evolve continuously from $0^\circ$ (collinear) to $90^\circ$ (coplanar) as a function of the coupling ratio $J_{aa}/|J_{ab}|$ (see Fig.~\ref{f:canting-angles}): the classical ground state has two continuous transitions.

Now, we observe that each of these hexamer states can tile the kagome
lattice.  Therefore, the ground state energy of each hexamer can be
used to deduce the ground state energy of the entire system.  In the
collinear regime ($J_{aa} > -|J_{ab}|/2$), the collinear hexamer
states lead to a unique global spin configuration (up to a global
SU(2) rotation), so there is long-range ferromagnetic order (if
$J_{ab}>0$) or ferrimagnetic order (if $J_{ab}<0$), and there is no
macroscopic residual entropy.  The ground state energy of the system
is
	\begin{equation}
	E_{\rm collinear} = {1 \over 6}(|J_{aa}| - 2 |J_{ab}|)N_{sites}~.
	 \label{e:collinear}
	\end{equation}
In the coplanar regime ($J_{aa}<-|J_{ab}|$), there are infinitely many
ways to tile the TKL with coplanar hexamer configurations (e.g.,
corresponding to 3-sublattice or 9-sublattice N\'eel order).
Furthermore, there are an infinite number of zero modes (rotations of
a few spins that cost zero energy).  The ground state energy is
	\begin{equation}
	E_{\rm coplanar} = -{1 \over 12} (|J_{aa}|+2 |J_{ab}|)N_{sites}~.
	 \label{e:coplanar}
	\end{equation}
The physics is essentially the same as that of the classical
Heisenberg kagome model.  For that model, the prevailing point of
view\cite{chalker1992,moessnerchalker1998,reimers1993,ritchey1993} is
that \emph{globally} coplanar configurations are selected at finite
temperature via an order-by-disorder mechanism, and the spin
chiralities develop nematic order; recently,
Zhitomirsky\cite{zhitomirsky2008} has argued that there is an
additional octupolar ordering which is, in fact, the true
symmetry-breaking order parameter.

The canted regime $-|J_{ab}| < J_{aa} < -|J_{ab}|/2$ has the
interesting property that in general $\alpha \ne \beta$, so there is a
net magnetic moment on each hexamer.  We have found that there are
still infinitely many ways to tile the TKL, and that there are still
an infinite number of zero modes.  It is possible that the zero modes
cause the directions f the local moment to vary from place to place,
destroying the long-range order with net magnetization; however, it is
conceivable that the spin correlation length gradually increases
towards infinity in going from the locally coplanar state to the
collinear state.  The energy of the canted state is
	\begin{align}
	E_{\rm canted} &= {2 \over 9} \bigg( -{7 |J_{aa}|\over 4} + {5 J_{ab}^2 \over 8 |J_{aa}|}  
	 \nonumber \\&
	 -|J_{ab}| \sqrt{\left(1 - {J_{aa}^2 / J_{ab}^2 }\right) \left({{J_{ab}}^2 / J_{aa}^2} - 1 \right) }	\bigg)N_{sites}.
	 \label{e:canted}
	\end{align}

Equations \eqref{e:collinear}, \eqref{e:coplanar}, and
\eqref{e:canted} are the exact ground state energies for the classical
Heisenberg model on the TKL.  They serve as upper bounds on the ground
state energy for the \emph{quantum} Heisenberg model.  Figure
\ref{f:energies} shows these upper bounds,
plotted together with the upper bounds derived from dimer
coverings, Eq.~\eqref{e:close-packed-vb-energy} and \eqref{e:dilute-vb}, as explained earlier in this section.  Notice that the
upper bound for the ground state energy set by considering dimer
configurations beats the classical ground states for $J_{aa}$ large
and negative (antiferromagnetic).  In this highly frustrated regime,
we expect that the true ground state of the quantum Heisenberg model
is significantly modified by quantum fluctuations from that of the
classical case.


\section{Conclusions \label{conclusions}}

In conclusion, we have studied the close-packed
dimer model on the triangular kagome lattice (TKL),
using exact analytic methods.
We find that (in the absence of an orienting potential) the entropy is
$s=\frac{1}{3} \ln 2$ per site, regardless of the weights of the
bonds, $z_{aa}$ and $z_{ab}$.  The occupation probability of every
bond is $p_\alpha=\frac{1}{4}$.  The dimer-dimer correlation function
vanishes identically beyond two lattice sites, faster than that in the
triangular lattice, and similar to the falloff in the case of the
kagome lattice.\cite{misguich-2002}
The monomer-monomer correlation function is $M(r)=1/4$
for $r$ greater than two lattice constants, indicating that
monomers are deconfined in this lattice.  This implies
that the Rokhsar-Kivelson point of the corresponding quantum dimer model
is a short-ranged, deconfined spin liquid.

In addition, we find that the classical ground state
of the Heisenberg model on the TKL is ferromagnetic (if $J_{ab}$ is ferromagnetic) or ferrimagnetic (if $J_{ab}$ is antiferromagnetic) when the coupling between $a$ spins on small trimers is large enough
compared to the coupling between $a$ spins and $b$ spins, $J_{aa} > -|J_{ab}|/2$. 
For $J_{aa}<-|J_{ab}|$, the ground state of a single hexamer is a coplanar state, and the physics reduces to that of the classical Heisenberg kagome model.\cite{chalker1992,moessnerchalker1998,reimers1993,ritchey1993}  
In between, there is a \emph{canted} classical
ground state in which the $a$ spins and $b$ spins within a hexamer both cant away from the coplanar state. 
Such a state does not arise in a simple model of
frustrated magnetism on the kagome lattice.
This type of canted ground state of the hexamer can tile the
lattice, and therefore it is the building block of the classical ground state
of the macroscopic system.   There is a corresponding macroscopic degeneracy associated with the many ways in which this local hexamer ground state can tile the lattice.  Each hexamer possesses a local moment; it is not yet clear whether the local magnetic moments from different hexamers cancel out due to the presence of zero modes.

\section*{Acknowledgments} 
It is a pleasure to thank M.~Ma for helpful discussions.
D.~X.~Y. acknowledges support from Purdue University.
This work was also supported by Research Corporation (Y.~L.~L. and E.~W.~C.).



\end{document}